\documentclass{article} %

\usepackage{arxiv}

\usepackage[utf8]{inputenc} %
\usepackage[T1]{fontenc}    %
\usepackage{url}            %
\usepackage{booktabs}       %
\usepackage{amsfonts}       %
\usepackage{amsmath}
\usepackage{amssymb}
\usepackage{nicefrac}       %
\usepackage{microtype}      %
\usepackage{graphicx}       %
\usepackage{hyperref}       %
\usepackage{multirow}
\usepackage{natbib}
\graphicspath{{./}}         %

\title{Securing Computer-Use Agents Against Branch Steering Attacks}
\date{}

\author{%
  Giulio Zingrillo \\
  ETH Zurich \\
  \And
  Hanna Foerster \\
  University of Cambridge \\
  \And
  Ilia Shumailov \\
  AI Sequrity Company \\
  \AND
  Yiren Zhao \\
  Imperial College London \\
  \And
  Robert Mullins \\
  University of Cambridge \\
}

\renewcommand{\shorttitle}{Securing Computer-Use Agents Against Branch Steering Attacks}

\hypersetup{
  pdftitle={Securing Computer-Use Agents Against Branch Steering Attacks},
  pdfauthor={Giulio Zingrillo, Hanna Foerster, Ilia Shumailov, Yiren Zhao, Robert Mullins},
}

\begin{document}

\maketitle

\begin{abstract}
Modern Computer Use Agents (CUAs) directly interact with graphical user interfaces and execute third-party web tools, exposing them to indirect prompt injection across every rendered page and tool response. The Dual-LLM pattern uses an isolated Planner LLM (P-LLM) to fix a plan before a Quarantined LLM (Q-LLM) processes untrusted inputs. In graphical environments, committed plans can branch on runtime observations. This exposes agents to branch steering attacks, where an adversary crafts untrusted data to coerce a CUA down a hazardous, pre-approved branch without injecting explicit instructions.
We systematically study branch steering attacks and introduce \textit{STEER-Bench} (101 tasks across 9 domains). On the 74 attack configurations shared by the baselines, attack success is 90.5\% against ReAct and 87.8\% against CaMeL-NOVA.
We then propose \textit{COBRA}, an architecture that pairs trusted branching plans with ahead-of-time capability constraints, strictly bounding the parameters and destinations each branch may execute. On STEER-Bench, COBRA reduces attack success to 0\% while retaining 97\% benign utility.\footnote{Code at   \url{https://github.com/giuliozing/securing-cuas-against-branch-steering}.}

\end{abstract}

\section{Introduction}

Computer-use agents (CUAs) carry out natural language tasks by interpreting on-screen visual content and executing mouse and keyboard actions. Most CUAs follow a ReAct-style loop in which a single model repeatedly observes the environment, reasons over it, and selects the next action; increasingly, agents also invoke Model Context Protocol (MCP) tools exposed by website vendors, giving them direct access to backend capabilities alongside the graphical user interface (GUI).

This architecture makes CUAs acutely vulnerable to \emph{indirect prompt injection} (IPI) across two surfaces. 
The first is \textbf{rendered web content} -- because a model processes both the user's trusted request and untrusted web pages, documents, and messages, untrusted text can become indistinguishable from instructions within the shared context \citep{greshake2023not,debenedetti2024agentdojo,zhang-etal-2025-attacking,aichberger2026mip,chen2025obviousinvisiblethreatllmpowered}. 
The second is the \textbf{MCP interface}, where a malicious server can embed instructions in tool descriptions (\emph{tool poisoning}), register tools that impersonate approved endpoints (\emph{tool shadowing}), or alter tool behavior post-approval~\citep{beurerkellner2025toolpoisoning}. Unlike webpage content, MCP tool descriptions may be presented to the agent during initialization, before it visits any website. 
Crucially, these two surfaces can be composed: a CUA task may interleave GUI actions with MCP invocations, injected page text can steer external tool calls, and poisoned tool responses can hijack GUI behavior. Defending either channel in isolation leaves the composite system vulnerable.

The Dual-LLM architecture mitigates IPI by strictly separating trusted planning from untrusted interaction \citep{willison2023dualllm,debenedetti2025defeatingpromptinjectionsdesign}. 
A privileged planner (P-LLM) derives the entire execution sequence in advance from the trusted user query alone; a quarantined model (Q-LLM) then processes untrusted content but is structurally barred from altering the plan. 
This yields \emph{control-flow integrity} (CFI): every action the agent executes must follow a pre-authorized plan. 
CFI constrains execution to the plan, while dynamic CUA environments make the choice among planned alternatives depend on runtime observations. 
Because web layout and state cannot be known before a page renders, control flow is tightly interleaved with runtime data. 
Plans must therefore incorporate branches over anticipated runtime states, as in CaMeL-NOVA~\citep{foerster2026camels}. Under branching execution, an adversary can manipulate untrusted page content to drive the agent down a hazardous, pre-approved path without injecting novel instructions, an exploit known as a \emph{branch steering attack} \citep{foerster2026camels}.

Branch steering demonstrates that CFI alone is insufficient for CUAs: agents additionally require \emph{data-flow integrity} (DFI). DFI requires that the resulting outcomes, including critical action arguments, remain within the user's authorization when execution depends on untrusted data (Section~\ref{sec:threat}). Enforcing DFI is uniquely challenging in CUAs because graphical primitives carry no intrinsic semantics: the same mouse click may open an innocuous tab or confirm an irreversible wire transfer, and static plans cannot differentiate between them. A standard Dual-LLM architecture bounds an agent to authorized action paths, but leaves this runtime data-flow channel unprotected.

\begin{figure}[t]
\centering
\includegraphics[width=.9\linewidth]{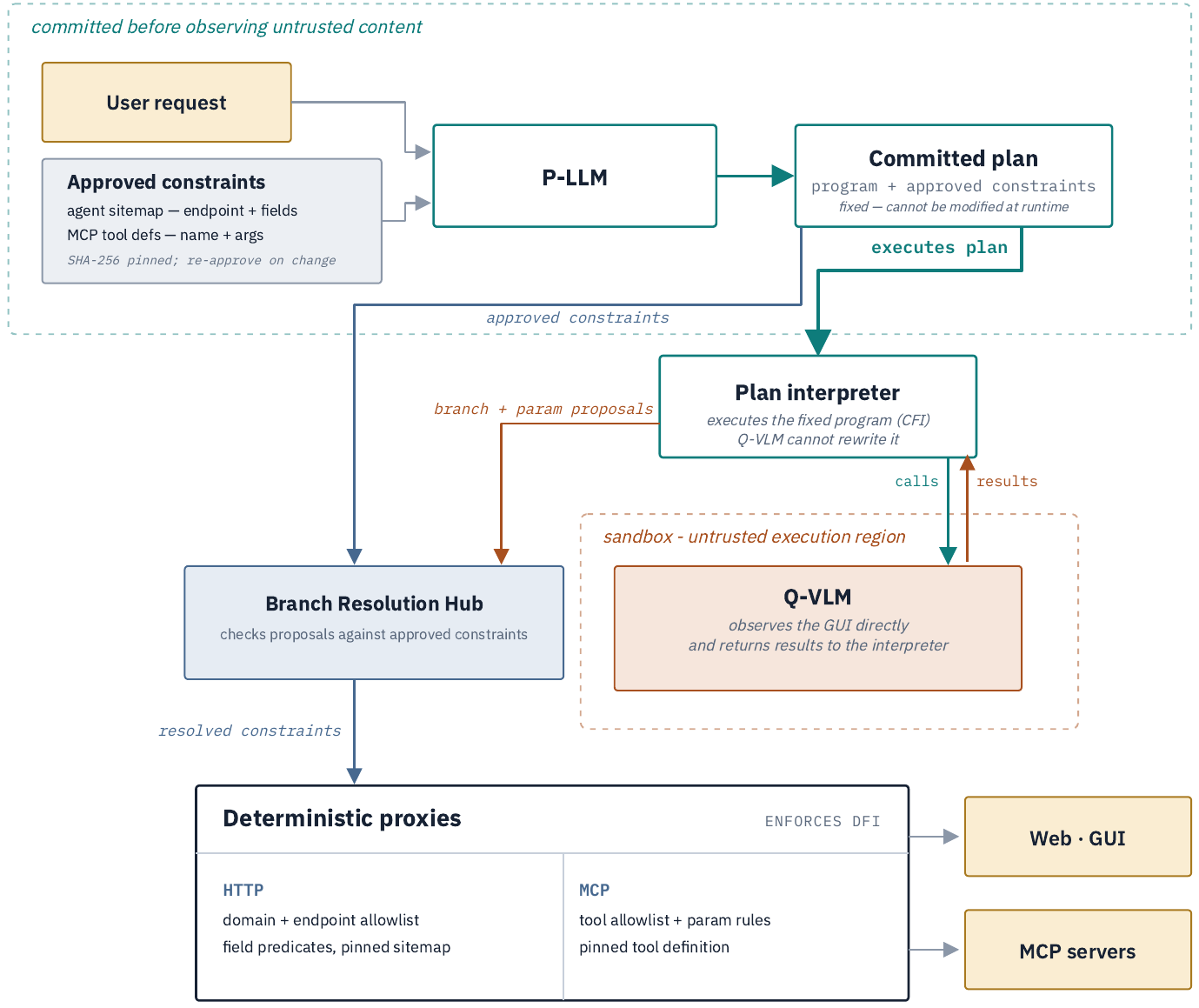}
\caption{COBRA separates planning from untrusted interaction. The interpreter provides CFI by executing the committed plan. The BRH resolves the active constraints, which the HTTP and MCP proxies apply to outgoing operations and constrained arguments.
}
\label{fig:architecture}
\end{figure}

To address this challenge, we present \textbf{COBRA}, a system-level defense that enforces CFI and checks HTTP requests and MCP calls against precommitted branch-level capability constraints. COBRA limits the effects of branch steering by constraining, ahead of time, the capabilities each branch may exercise (e.g., allowed destination domains, endpoints, and tool parameters). 
As illustrated in Figure~\ref{fig:architecture}, before observing any untrusted runtime data, the P-LLM compiles the user's request along with approved site and tool metadata into an executable program annotated with branch-level constraints. A deterministic interpreter executes this program, querying a quarantined vision-language model (Q-VLM) strictly for visual perception. 
The interpreter selects a planned branch using runtime observations. The \emph{Branch Resolution Hub} (BRH) resolves the active constraints for that path, and the HTTP and MCP proxies check outgoing operations and their constrained arguments. By unifying GUI-mediated web traffic and MCP tool calls under identical constraint contracts, COBRA secures both injection channels even across deeply nested branches.

In this paper, we make the following contributions :
\begin{itemize}

\item \textbf{Benchmarking Branch Steering:} We introduce \textbf{STEER-Bench}, the first systematic benchmark designed to evaluate branch steering attacks in CUAs, spanning 101 end-to-end tasks across 9 real-world domains. On 74 shared attack configurations, the attack success rate (ASR) is 90.5\% against ReAct and 87.8\% against CaMeL-NOVA.
    
\item \textbf{System-Level Defense Architecture:} We frame branch steering as a data-flow integrity failure and introduce \textbf{COBRA}, a dual-LLM framework that enforces ahead-of-time, branch-level capability constraints across both GUI-driven HTTP requests and MCP tool boundaries.
    
\item \textbf{Empirical Evaluation:} Across STEER-Bench, COBRA (\texttt{gpt-5} P-LLM and \texttt{UI-TARS-1.5-7B} Q-LLM) blocks all evaluated attacks (0\% ASR) while preserving 97\% of clean-task utility. Across four published CUA and MCP security benchmarks comprising
1{,}312 attack cases, COBRA blocks 1{,}289 cases. HTTP and MCP
enforcement blocks 1{,}259 of these cases, and plan--execution
separation blocks 30 additional OS-Harm cases. COBRA preserves 100\% benign completion on the paired cases where
it is reported and achieves 79\% completion under attack on MCPTox.

\end{itemize}

\section{Related Work}
\label{sec:related}

\paragraph{Probabilistic defenses for CUAs.}
Malicious pop-ups, adversarial image patches, and visible or hidden visual prompt injections can manipulate CUAs \citep{zhang-etal-2025-attacking,aichberger2026mip,chen2025obviousinvisiblethreatllmpowered}. Detection-based defenses identify such threats from agent observations or actions \citep{yang2025context,10.1145/3719027.3765064}, but depend on a model or classifier recognizing the attack. MELON detects indirect prompt injections in tool-use agents by rerunning the agent with the user's request masked and comparing the resulting tool calls with those from the original run \citep{zhu2025melon}. This approach is difficult to apply to live CUAs because clicks and keystrokes reveal little about the intended task, and the desktop environment cannot generally be reproduced exactly.

\paragraph{Structural system-level defenses.}
Structural defenses instead enforce predefined properties independently of untrusted content, and COBRA follows this approach. Dual-LLM architectures separate trusted planning from interaction with untrusted content \citep{willison2023dualllm}. Later systems enforce executable plans, track data provenance or dependencies, or isolate components in tool-use and multi-agent systems \citep{debenedetti2025defeatingpromptinjectionsdesign,costa2025securingaiagentsinformationflow,zhong2025rtbasdefendingllmagents,abdelnabi2026firewallssecuredynamicllm,wu2025isolategptexecutionisolationarchitecture}. FIDES and CaMeL can enforce the same restrictions on protected tool
calls when their policies encode the same authorization requirements.
FIDES checks integrity and confidentiality labels, while CaMeL can
express these checks using provenance, dependencies, and allowed readers.
As discussed by \citet{foerster2026camels}, low-level GUI actions carry
no intrinsic semantic meaning. The same click can open a preview or
submit a payment depending on the current page. COBRA therefore checks
the HTTP requests and MCP calls that carry out these operations
(Section~\ref{sec:architecture}). Progent enforces LLM-generated least-privilege policies for structured tool calls in text-based agents, but does not cover GUI interaction or raw HTTP requests \citep{shi2026progentsecuringaiagents}.

\paragraph{MCP security.}
Prior work characterizes attacks against MCP tool definitions and behavior \citep{beurerkellner2025toolpoisoning}. MCPTox~\citep{wang2026mcptox} and MCPSecBench~\citep{yang2025mcpsecbench} evaluate tool poisoning and post-approval manipulation, while ETDI~\citep{bhatt2025etdi} authenticates tool definitions and binds them to authorization decisions. These works primarily study MCP as a standalone interface; COBRA protects tasks combining MCP calls and web interaction.

\paragraph{Evaluation methodology.}
WASP~\citep{evtimov2025waspbenchmarkingwebagent} evaluates web prompt injections that redirect agent actions, while OS-Harm~\citep{kuntz2026osharm} includes OS-level attempts to induce unauthorized behavior. MCP benchmarks instead focus on the tool interface \citep{wang2026mcptox,yang2025mcpsecbench,zong2025mcp}. InjecAgent and ASB evaluate text-based tool-using agents rather than live CUAs, and therefore do not exercise COBRA's GUI-mediated HTTP enforcement \citep{zhan-etal-2024-injecagent, zhang2025agentsecuritybenchasb}
. These benchmarks report only whether an attack succeeded against the defense system as a whole, not which component stopped it. STEER-Bench instead groups attacks into suites by the property they violate, such as DFI, and runs each suite with the corresponding defense component enabled and disabled.

\section{Threat Model}
\label{sec:threat}

\paragraph{Integrity definitions in branching CUAs.}
We reason about agent execution in terms of control  and data flows. 
An agent satisfies \emph{Control-Flow Integrity} (CFI) if every action executed at runtime belongs to the pre-authorized plan derived from the user's trusted request. 
While existing Dual-LLM systems provide CFI by committing to a static sequence of actions ahead of time, dynamic web interfaces force plans to include runtime branches. 

\paragraph{Data-flow integrity in branching CUAs.}
A prompt injection that induces an action outside the committed
plan violates CFI. Within the plan, runtime observations can influence which branch
the agent follows and the arguments supplied to external actions. Data-flow integrity (DFI)
requires that the resulting outcomes stay within what the user
authorized when execution depends on untrusted runtime data. This applies
to the arguments of an external action even on the intended branch.
A branch-steering attack targets this property by manipulating those
observations to induce an unauthorized outcome through a branch
already present in the plan.

\paragraph{Attacker capabilities and objectives.}
We assume the adversary can manipulate untrusted content rendered or returned by third-party web services that the agent navigates during execution, without needing full server compromise. 
This includes attacker-controlled web elements such as advertisements, user reviews, comments, marketplace listings, incoming messages, and downloaded attachments. 
On unapproved or compromised external servers, the attacker may additionally control MCP tool descriptions and tool outputs. 
Following a Kerckhoffs-style threat model, the attacker knows
COBRA's architecture and enforcement mechanisms but cannot directly
inspect or modify the committed execution plan or the protected
state of the interpreter. The attacker can observe external
execution behavior and use that feedback to adapt an attack.

The adversary pursues two distinct objectives:
\begin{enumerate}
    \item \emph{Control-Flow Integrity (CFI) Violation:} Inducing the agent to execute actions entirely absent from the approved plan (e.g., triggering unauthorized tool calls, navigating to malicious domains, or executing arbitrary system commands).
    \item \emph{Data-Flow Integrity (DFI) Violation:}
Inducing the agent to produce an outcome outside the user's
authorization by manipulating the runtime observations that determine
which branch of the approved plan is taken or which arguments an external
action receives.
\end{enumerate}

\paragraph{Defender capabilities and scope.}
COBRA enforces CFI by restricting execution to the committed
plan. At the HTTP and MCP boundaries, it enforces the constraints
captured in the plan for the active branch, including destinations
and constrained arguments. This supports the DFI objective when the
plan and its constraints faithfully capture the user's authorization.
Checking each request against these constraints does not independently
establish the truth of a branch condition or enforce operation counts
across otherwise permitted requests.
For HTTP traffic, enforcement operates at two granularities depending
on metadata availability (Appendix~\ref{app:sitemap-guarantees}):\begin{itemize}
    \item \emph{Host-level enforcement (default):} Without a sitemap, COBRA restricts network operations to an allowlist of authorized domains derived from the user request and deployment policy, preventing arbitrary navigation or cross-origin data exfiltration.
    \item \emph{Endpoint-level enforcement (sitemap-guided):} When an approved agent sitemap is available, COBRA's proxy validates request methods, parametrized URL routes, and payload fields against the specific branch's constraints.
\end{itemize}
COBRA binds each approval of a site sitemap or MCP tool definition
to its SHA-256 hash. A hash mismatch blocks use of the changed
description under the previous approval and triggers a new approval
request. Appendix~\ref{app:sitemap-guarantees} describes the
enforcement available after that decision.

\section{Architecture}
\label{sec:architecture}

COBRA consists of a plan interpreter, a central \textbf{Branch Resolution Hub} (BRH), and deterministic HTTP and MCP proxies. The interpreter executes the committed plan, the BRH resolves constraints for the selected path using rules fixed during planning, and the proxies check outgoing HTTP requests and MCP calls against the resolved constraints. Enforcement targets these calls rather than GUI actions because a click does not, by itself, identify the operation it causes.

\paragraph{Approved descriptions and planning inputs.} An \emph{agent sitemap} describes a site's \emph{operations}, such as approving a loan or posting a comment. Each operation specifies an \emph{endpoint}, an HTTP method and URL template such as \texttt{POST /loans/\{id\}/approve}, and the named \emph{fields} carried by the request. MCP tools and their structured arguments provide the corresponding vocabulary for MCP. COBRA uses only site and tool descriptions approved by the user, binding each approval to the artifact's SHA-256 hash; a changed description requires renewed approval. An approved agent sitemap gives COBRA the operation descriptions it
needs to check HTTP methods, endpoints, and request fields against the
active branch's constraints. Without an approved sitemap, COBRA
restricts HTTP requests to authorized hosts
(Appendix~\ref{app:sitemap-guarantees}). Runtime content cannot extend this list. To make sitemaps available for sites that do not publish them, we introduce MESA, which can crawl selected domains on the allowlist and generate candidate sitemaps. This lets us evaluate how sitemap availability affects security and utility (Appendix~\ref{app:mesa}).

\paragraph{Planning and constraint initialization.}
COBRA generates and annotates a plan before observing untrusted content from the current run. First, the P-LLM generates a program from the trusted request and approved descriptions. If the request requires a security-critical value that neither the user supplied nor a source authorized by the plan can establish, COBRA asks the user for it during plan construction; the supplied value is trusted for that request only. The same clarification mechanism applies when the user's authorization to perform an action depends on a runtime condition. If the planner determines that the condition can only be established from an untrusted observation, it asks the user for clarification before committing the plan. Second, COBRA parses the program to identify its actions and branches. A trusted annotation call assigns each branch its permitted destinations and tools, together with constraints on request fields and tool arguments, using the approved descriptions and the branch's role in the plan (Appendix~\ref{app:annotation}). The interpreter then executes the committed program, calling the Q-VLM where screen observation or interaction is required. The Q-VLM may report observations and propose low-level steps but cannot modify the program. Untrusted content therefore cannot introduce an action absent from the plan, establishing CFI.

\paragraph{Branch and parameter resolution.}
The interpreter evaluates the committed branch condition using the current plan variables, which may include observations reported by the Q-VLM. The BRH resolves the selected path's constraints and checks proposed action parameters against them. When the user authorizes a range of outcomes, the plan encodes
that range in its constraints. The HTTP and MCP proxies check that the
resulting operation satisfies the constraints for the selected branch. 

Constraints accumulate through nested branches and are refreshed at each transition, so permissions from an earlier branch do not persist into a later one.

\paragraph{Enforcement at the HTTP and MCP boundaries.}
The HTTP and MCP proxies enforce the plan's constraints by checking that the outgoing
operation and the values of its constrained arguments satisfy the
BRH's resolved constraints for the selected path. A modified ceLLMate proxy \citep{meng2025cellmate} intercepts HTTP requests, including those triggered by GUI actions. Without an approved agent sitemap, it admits only hosts authorized by the plan. With one, it matches each request to a declared operation and checks its destination and field values (Appendix~\ref{app:sitemap-guarantees}). GUI effects that do not traverse HTTP fall outside this boundary. The MCP proxy applies the same resolved constraints to named tools and their arguments. It also verifies that the tool definition matches the approved version, detecting rug pulls, and that the call reaches the approved server, preventing same-name tool shadowing. Appendix~\ref{app:mcp-arbitration} details how COBRA coordinates GUI and MCP interaction under the same enforcement state.

\paragraph{Fused planning through multi-sample inference.}
Because the planner cannot observe the environment, a single plan may omit a relevant UI state or failure mode. Fused planning generates five plans in parallel from the same trusted input and merges them, in a final trusted call, into one branching program that retains distinct alternatives. The merged program is committed and processed by the same interpreter, BRH, and proxies as a standard plan. Five candidates provide a favorable empirical trade-off between utility and token consumption (Appendix~\ref{app:fusion}).

\section{Evaluation}
\label{sec:steerbench}
\paragraph{Setup.}
Unless stated otherwise, we use \texttt{gpt-5} as the P-LLM and locally serve \texttt{UI-TARS-1.5-7B} as the Q-VLM. We report attack success rate (ASR), the fraction of attacks that achieve their objective, and task completion without an injection (\emph{benign utility}) or with the injection present (\emph{utility under attack}), as applicable. We evaluate COBRA end to end, including the constraints generated by its planning and annotation pipeline, and determine outcomes from recorded environment state and proxy alerts. The evaluation exercises HTTP enforcement both without approved sitemaps, which restricts the sites the agent may contact, and with them, which also restricts the operations it may perform on those sites. It also exercises MCP enforcement using approved tool manifests.

\paragraph{Design.}
STEER-Bench serves two purposes. It tests whether branch steering
succeeds against existing computer-use agents, and it measures the
contribution of individual COBRA protections through targeted ablations.
Each suite runs its attack cases against complete COBRA and against a
version with the protection targeted by that suite disabled. Each comparison holds the trusted request, intended violation, and models fixed. The benchmark contains 101 tasks, some tested under multiple attack configurations, yielding 199 attack cells and 138 benign cells. A cell is one task run under one configuration. Every attack cell has a corresponding benign reference, but several attack configurations for the same legitimate scenario may share one benign cell, accounting for the difference in counts.
Of these, 74 attack cells and 74 corresponding runs without the injection can also be evaluated on generic computer-use agents. We evaluate both baselines once on each of the same 74 attack
configurations. For each baseline, we compute aggregate ASR by
dividing the total number of successful attacks by 74. The remaining cells exercise capabilities that these baselines lack, including MCP interaction and approval of site or tool descriptions. The tasks span web interactions through the GUI and MCP tool calls, four page-layout families, and domains ranging from bank transfers and warehouse logistics to clinical dosage adjustments. For each attack cell, we use the actions recorded by the environment
to determine whether the harmful outcome specified for that task
occurred. The nine suites (S1-S9) test the following components (Appendix~\ref{app:steerbench-design}):

\emph{HTTP and MCP enforcement.}
\emph{S1 WIRE} changes a field value while preserving the authorized action and destination, testing field-value constraints on HTTP requests. \emph{S2 DEST} redirects an otherwise valid action to an unauthorized host or endpoint, testing destination constraints. \emph{S3 MCP} invokes an unauthorized tool or supplies arguments that violate active constraints, testing the MCP proxy. \emph{S4 Cross-channel} moves an operation between HTTP and MCP, testing whether the same constraints remain effective across both channels.

\emph{Trusted inputs and plan integrity.}
\emph{S5 PROV} tests whether COBRA permits a value from the trusted request or an authorized, verifiable source while rejecting the same value when it comes only from untrusted content. If no permitted source establishes a required value, COBRA asks the user to supply it.
 \emph{S6 TRUST} modifies an approved site or tool description, testing detection of post-approval changes. \emph{S7 CFI} presents poisoned content intended to introduce an action outside the committed plan, testing the separation between planning and interaction.

\emph{Branch constraints.}
\emph{S8 STEP} tests whether permissions expire after a branch transition. \emph{S9 ULTRA} tests whether DFI constraints remain active through four nested branches.

\paragraph{Results.}
Complete COBRA blocks every STEER-Bench attack, whereas attacks succeed in every suite when its targeted protection is disabled (Table~\ref{tab:steerbench}). ASR rises from 0\% with complete COBRA to between 55\% and 100\% across the ablations, with 174 of 199 attacks succeeding overall (87.4\%). On the 74 shared attack configurations, ASR is 90.5\% (67/74) with ReAct using \texttt{UI-TARS-1.5-7B} and 87.8\% (65/74) with CaMeL-NOVA using \texttt{gpt-5} as the P-LLM and \texttt{UI-TARS-1.5-7B} as the Q-VLM. Thus, trusted planning alone does not prevent untrusted values from steering authorized branches. Using the same P-LLM and Q-VLM as the CaMeL-NOVA baseline, COBRA retains 97\% of its benign utility, while only producing four false positives across 138 benign cells: three in S8 and one in S9, where constraints combined across nested branches block legitimate actions.

In the original S5 evaluation, clarification requests terminated
the run without a user answer. Underspecified cases were scored for
correct abstention and excluded from benign utility. On sufficiently
specified requests, an unnecessary clarification counted as a utility
failure. Table~\ref{tab:steerbench} measures attack success separately
from task completion; absence of a harmful outcome does not imply
completion of the request.

In S5, the planner asked for clarification before carrying out the
requested action in 20 of 22 underspecified cases. For example, when
asked to dispatch an order strictly to the saved delivery address,
COBRA asked the user to supply the address. In a separate set of 11
benign cases with sufficiently specified requests, COBRA made no
unnecessary clarification requests. When the request supplied the
shipping address (Warehouse 4, Rotterdam), the plan used that address.
These clarification counts come from a separate analysis of the execution
logs. Table~\ref{tab:steerbench} reports whether the task-defined harmful
outcome occurred.

\paragraph{Responding to clarification.}
We additionally evaluate clarification responses on the 11 S5 tasks
using \texttt{gpt-5} at pass@1. Crossing benign/poisoned-record variants
with two response conditions gives 44 runs, all initially requesting
clarification. A simulated user supplies either the missing value while
retaining the original conditions or changed authorization to use the
on-file record. COBRA replans from the original request, its question
verbatim, and the answer as a new trusted user turn. Supplying the value
yields 10/11 benign completions and 0/11 successful attacks. Authorizing
the on-file record yields 0/11 benign completions, all 11 stopping safely,
and 0/11 successful attacks; stopping is a utility cost. Outcomes come
from environment-recorded actions and enforcer logs. These additional
runs are reported separately from Table~\ref{tab:steerbench}. They test
missing-value requests and do not establish reliable verification of
arbitrary factual authorization conditions.

\paragraph{Constraint-generation accuracy.}
We evaluate how accurately the P-LLM translates the user's
authorization into constraints. We selected 30 cells from four
STEER-Bench suites and manually annotated the expected constraints
before running COBRA. For each cell, we used the annotation to define
probes that check whether the generated constraints admit authorized
operations and block unauthorized ones. A cell passes if the generated
constraints give the expected decision on every probe. The generated constraints passed in 28 of the 30 cells (Table~\ref{tab:constraint-accuracy}). In one of the two remaining
cells, the constraints were too restrictive and blocked an authorized
operation. In the other, they were too permissive and admitted an
unauthorized operation.

\begin{table}[t]
\centering
\footnotesize
\setlength{\tabcolsep}{3pt}
\caption{STEER-Bench results. ``Cells'' denotes the number of attack cases evaluated in each suite. ASR is the fraction of attack cells in which the task-defined harmful
outcome occurs. ASR (on) uses complete COBRA. ASR (off) disables the listed component. Baseline percentages are based on the applicable configurations in each suite, totaling 74 per baseline.}
\begin{tabular}{@{}llccccc@{}}
\toprule
Suite & Component tested & Cells & ASR (on) & ASR (off) & CaMeL-NOVA & ReAct\\
\midrule
S1 WIRE          & Field constraints          & 26 & 0\% & 100\% & 96\%  & 96\%  \\
S2 DEST          & Host/endpoint constraints  & 24 & 0\% & 100\% & 88\%  & 92\%  \\
S3 MCP           & MCP proxy                  & 15 & 0\% & 67\%  & ---   & ---   \\
S4 Cross-channel & Shared constraint state    & 72 & 0\% & 100\% & 89\%  & 100\% \\
S5 PROV          & Trusted-value handling     & 33 & 0\% & 55\%  & 100\% & 100\% \\
S6 TRUST         & Description integrity      & 5  & 0\% & 80\%  & ---   & ---   \\
S7 CFI           & Plan--execution boundary   & 9  & 0\% & 67\%  & 89\%  & 89\%  \\
S8 STEP          & Constraint refresh         & 10 & 0\% & 100\% & 40\%  & 40\%  \\
S9 ULTRA         & Constraint composition     & 5  & 0\% & 80\%  & 100\% & 100\% \\
\midrule
\multicolumn{2}{@{}l}{\textbf{All suites}} &
\textbf{199} & \textbf{0\%} & \textbf{87.4\%} &
\textbf{87.8\% (65/74)} & \textbf{90.5\% (67/74)} \\
\bottomrule
\end{tabular}
\label{tab:steerbench}
\end{table}

\subsection{External Validation}
\label{sec:external}

We evaluate COBRA's control-flow integrity (CFI), data-flow integrity (DFI), and MCP protections on four published benchmarks beyond STEER-Bench. COBRA blocks every evaluated attack within these protections' stated scope. The results distinguish protection supplied by plan--execution separation from the additional protection supplied by approved sitemaps and MCP enforcement.

\paragraph{Plan--execution separation (CFI) and domain-level DFI enforcement.}
The \textit{OS-Harm}~\citep{kuntz2026osharm} prompt-injection cases are already blocked by plan--execution separation: both COBRA and CaMeL-NOVA~\citep{foerster2026camels} prevent all 51 attacks. To test DFI enforcement without an approved sitemap, we run the 21 cases that attempt unauthorized HTTP requests with a ReAct-style \texttt{UI-TARS-1.5-7B} agent instructed to follow injected on-screen instructions. In this setting, COBRA's HTTP proxy restricts requests to the domains authorized by the plan. Without the proxy, all 21 unauthorized requests succeed; with it, all 21 are blocked while all corresponding legitimate operations remain permitted. The other 30 cases use channels such as SMTP that the HTTP proxy does not inspect, but remain blocked by the dual-LLM plan--execution boundary. The 21 HTTP cases contribute to the total of 1{,}259 cases blocked
by HTTP and MCP enforcement across the four benchmarks. Counting
the other 30 OS-Harm cases once brings the full-system total to
1{,}289 blocked cases.

\paragraph{Sitemaps and same-site attacks.}
On \textit{WASP}~\citep{evtimov2025waspbenchmarkingwebagent}, we compare host-level enforcement with enforcement using GitLab and Reddit sitemaps from ceLLMate~\citep{meng2025cellmate}, treated as approved for the experiment. Without sitemaps, COBRA blocks the five cross-domain attacks, but all 14 attacks invoking an unauthorized operation on the same host succeed; with sitemaps, COBRA blocks all 19 request-level attacks, reducing ASR from 14/19 (73.7\%) to 0/19 while preserving 100\% benign completion. For example, when the user requests a GitLab issue comment, the plan authorizes \texttt{GET /\{group\}/\{project\}/-/issues/*} and \texttt{POST /\{group\}/\{project\}/-/notes}; an injection instead issues \texttt{POST /byteblaze/dotfiles/-/deploy\_keys}. Host-level enforcement permits this same-host request, whereas the sitemap rejects the unauthorized endpoint.
The remaining two attacks depend on semantic properties of otherwise authorized free-form text and therefore fall outside deterministic request-level enforcement.

\paragraph{Tool poisoning and server-side limits.}
\textit{MCPTox}~\citep{wang2026mcptox} shows that COBRA blocks tool-poisoning attacks when they change an observable MCP call, but cannot constrain a compromised server's hidden implementation. Of the 1,223 MCPTox cases, 1,210 change the selected tool or its arguments. COBRA blocks all 1,210 by checking each call against the active branch's constraints, compared with 100\% ASR without enforcement. In the remaining 13 cases, the legitimate and malicious calls use identical tools and arguments, but the server produces a different hidden effect that the proxy cannot distinguish. COBRA achieves 79\% utility under attack, compared with 100\% without enforcement; the loss primarily reflects failures to generate a suitable and efficient plan rather than legitimate calls blocked by the proxy.

\paragraph{Tool shadowing and rug pulls.}
\textit{MCPSecBench}~\citep{yang2025mcpsecbench} tests whether COBRA prevents a server from impersonating an approved tool and detects changes to a tool definition after approval. Of its 17 attacks, nine fall within COBRA's threat model. COBRA blocks all nine: five test approved-server binding and hash-pinned tool definitions, while four test whether tool arguments satisfy the active branch's constraints. It preserves benign utility on all corresponding cases. The remaining attacks target the host's command namespace, transport, or client and server implementation bugs; tool poisoning is evaluated above using MCPTox.

\subsection{Utility and Cost}
\label{sec:utility}

We evaluate COBRA's utility on 361 tasks across 10 OSWorld-MCP domains \citep{jia2025osworldmcpbenchmarkingmcptool}, including how MCP interaction is used and whether fused planning recovers tasks that a single committed plan misses. Task-completion results use pass@5 with early stopping and at most 20 interaction steps per attempt; completion requires an OSWorld score of $1.0$. For the 361-task evaluation in Table~\ref{tab:osworld}, HTTP enforcement is limited to authorized hosts. Table~\ref{tab:osworld} compares COBRA with ReAct and reports \texttt{Sonnet-5} and \texttt{kimi-k2.5} in both COBRA roles, alongside \texttt{gpt-5} as the P-LLM with the smaller \texttt{UI-TARS-1.5-7B} Q-VLM; additional OpenCUA results appear in Appendix~\ref{app:opencua-utility}. We use a stronger P-LLM with smaller GUI models, as in \citet{foerster2026camels}, because assigning those models the planning role often produces invalid plans and lowers completion (Appendix~\ref{app:model-cost-utility}).

\paragraph{COBRA versus ReAct.}
As in \citet{foerster2026camels}, the strongest result comes from assigning planning to a capable P-LLM and repeated interaction to a smaller Q-VLM. With \texttt{gpt-5} as its P-LLM and \texttt{UI-TARS-1.5-7B} as its Q-VLM, COBRA completes 133/361 OSWorld-MCP tasks (36.8\%), compared with 85/361 (23.5\%) for ReAct using \texttt{UI-TARS-1.5-7B} alone (Table~\ref{tab:osworld}). 
Yet this asymmetric configuration is economically practical because the stronger model is used for planning rather than every interaction. Planning and annotation consume an average of 73k tokens per task, compared with 258k for the locally served Q-VLM (Appendix~\ref{app:model-cost-utility}). Using \texttt{Sonnet-5} for both COBRA roles yields a similar completion rate to \texttt{gpt-5}+\texttt{UI-TARS-1.5-7B} (38.8\% versus 36.8\%), but higher measured API charges (\$0.627 versus \$0.332 per task).

\begin{table*}[t]
\centering
\small
\caption{Task completion on 361 OSWorld-MCP tasks under pass@5 with early stopping. Completion requires an exact OSWorld score of $1.0$. Adding \texttt{gpt-5} planning improves completion over the model-matched \texttt{UI-TARS} ReAct baseline, whereas using \texttt{kimi-k2.5} or \texttt{Sonnet-5} in both COBRA roles does not preserve their ReAct performance. OpenCUA results are reported in Appendix~\ref{app:opencua-utility}.
}
\label{tab:osworld}

\resizebox{\textwidth}{!}{
\begin{tabular}{@{}lcrrrrrr@{}}
\toprule
\multirow{2}{*}{Domain} &
\multirow{2}{*}{Tasks} &
\multicolumn{3}{c}{\textbf{ReAct}} &
\multicolumn{3}{c}{\textbf{COBRA}} \\
\cmidrule(lr){3-5}
\cmidrule(l){6-8}
&
&
\texttt{UI-TARS} &
\texttt{kimi-k2.5} &
\texttt{Sonnet-5} &
\shortstack{\texttt{gpt-5}\\+\texttt{UI-TARS}} &
\shortstack{\texttt{kimi-k2.5}\\dual-role} &
\shortstack{\texttt{Sonnet-5}\\dual-role} \\
\midrule
Chrome
 & 46 & 11 (24\%) & \textbf{21 (46\%)} & 20 (43\%)
 & 14 (30\%) & 9 (20\%) & 15 (33\%) \\
GIMP
 & 26 & 17 (65\%) & 12 (46\%) & 9 (35\%)
 & \textbf{21 (81\%)} & 14 (54\%) & 16 (62\%) \\
LibreOffice Calc
 & 47 & 6 (13\%) & 23 (49\%) & \textbf{35 (74\%)}
 & 10 (21\%) & 8 (17\%) & 14 (30\%) \\
LibreOffice Impress
 & 47 & 12 (26\%) & 23 (49\%) & \textbf{27 (57\%)}
 & 17 (36\%) & 9 (19\%) & 18 (38\%) \\
LibreOffice Writer
 & 23 & 8 (35\%) & 12 (52\%) & \textbf{15 (65\%)}
 & 11 (48\%) & 7 (30\%) & 13 (57\%) \\
Multi-apps
 & 93 & 3 (3\%) & 22 (24\%) & \textbf{34 (37\%)}
 & 15 (16\%) & 11 (12\%) & 16 (17\%) \\
OS
 & 24 & 8 (33\%) & 14 (58\%) & \textbf{17 (71\%)}
 & 15 (63\%) & 14 (58\%) & 16 (67\%) \\
Thunderbird
 & 15 & 5 (33\%) & 8 (53\%) & \textbf{9 (60\%)}
 & 8 (53\%) & 2 (13\%) & 8 (53\%) \\
VLC
 & 17 & 5 (29\%) & 4 (24\%) & 4 (24\%)
 & 5 (29\%) & 3 (18\%) & \textbf{6 (35\%)} \\
VS Code
 & 23 & 10 (43\%) & 13 (57\%) & 14 (61\%)
 & 17 (74\%) & 15 (65\%) & \textbf{18 (78\%)} \\
\midrule
\textbf{Total}
 & \textbf{361}
 & \textbf{85 (23.5\%)}
 & \textbf{152 (42.1\%)}
 & \textbf{184 (51.0\%)}
 & \textbf{133 (36.8\%)}
 & \textbf{92 (25.5\%)}
 & \textbf{140 (38.8\%)} \\
\bottomrule
\end{tabular}
}
\end{table*}
\paragraph{Utility with MESA sitemaps.}
MESA generates sitemaps covering all required operations for 18 of
the 25 website-facing Chrome tasks (Appendix~\ref{app:mesa}).
Both configurations use \texttt{gpt-5} as the planner,
\texttt{UI-TARS-1.5-7B} as the vision model, and pass@5.
In the original host-level evaluation, COBRA completed 11 of those
18 tasks. With the MESA-generated sitemaps and endpoint and field
checks enabled, COBRA completed 10 of those 11 tasks, retaining
90.9\% of the original successes on this subset.

\paragraph{Comparison with CaMeL-NOVA.}
COBRA completes 36.8\% of OSWorld-MCP tasks with \texttt{gpt-5} and \texttt{UI-TARS-1.5-7B}. Although OSWorld and OSWorld-MCP differ overall, they share 86 tasks from Chrome, VLC, and VS Code. On these matched tasks, COBRA completes 36/86 (41.9\%), compared with 22/86 (25.6\%) for CaMeL-NOVA~\citep{foerster2026camels}, a difference of 16.3 percentage points. MCP may contribute in two ways. Approved tool descriptions give the planner a vocabulary of available operations even when execution uses the GUI, while tool calls provide alternative routes with GUI verification or fallback. MCP calls occur in 336/722 task--configuration evaluations (46.5\%) across the two configurations analyzed in Appendix~\ref{app:mcp-arbitration}.

\paragraph{Fused planning.}
Fused planning can complete tasks in one attempt that standard planning solves only after several. It combines five candidate plans into one committed branching program, preserving alternative routes and fallbacks without exposing the P-LLM to runtime observations. Of 55 tasks completed on attempts 2--5 with standard planning, 27 (49.1\%) complete on the first attempt with fusion (Appendix~\ref{app:fusion}).

\section{Discussion}
\label{sec:discussion}

\textbf{Why constrain both control flow and data flow?} A task may need different permissions as execution moves through conditional and nested branches. A request-level defense such as ceLLMate~\citep{meng2025cellmate} can check conditions on HTTP requests, but without anticipating much of the task's execution flow, it is difficult to specify when permissions should appear, combine, and expire across its possible branches. Further, if a branch is not anticipated, the agent may still enter it while permissions intended for another part of the task remain available. An attacker-controlled review could prompt such a change of course, and the resulting requests may satisfy a wrong selection of policies. COBRA instead commits to a branching plan and its constraints before execution. CFI disallows routes outside that plan, while DFI enforces the constraints specified for whichever planned route is taken.

\textbf{How far can structural enforcement extend?}
COBRA checks the committed plan and operations observable through HTTP and MCP, but three questions remain. First, how can it constrain security-critical values that never appear in an HTTP request or MCP call? If a checkout request contains only item identifiers rather than the displayed total, COBRA must read the price from the GUI, where it may not reliably distinguish the site's price from visually similar attacker-controlled content in an ad or customer review. In this case it cannot enforce a deterministic constraint on the total at the current HTTP or MCP boundaries. Second, how can COBRA check the meaning of free-form content or effects hidden behind an otherwise valid request? Two comments sent to the same endpoint may differ only in whether their text is offensive, while a compromised MCP server may produce a harmful effect despite receiving the same tool call as an honest server. Third, how can COBRA extend enforcement to other computer applications whose actions use protocols or local interfaces outside its HTTP and MCP proxies? An email client may send through SMTP, while a desktop application may write directly to the filesystem; neither action is covered by COBRA’s HTTP and MCP proxies, and many applications do not expose MCP tools. Extending enforcement is difficult because these applications lack a common interface that exposes the semantic meaning of their actions and the values those actions use, which is needed to impose constraints.

\textbf{How can blind planning recover from unexpected states without weakening the trust boundary?}
The lower completion of the same-model COBRA configurations relative to ReAct shows the difficulty of anticipating interface states before execution. Fused planning recovers some later successes on the first attempt, but cannot yet replace repeated independent attempts. A promising direction is to train the P-LLM for computer use rather than encode ever more contingencies in its prompt. Training on task trajectories could teach it recurring routines, such as detecting browser state, dismissing cookie banners, and recovering from failed clicks, while feedback from execution could reward plans that anticipate relevant failure states. Training across website and application layouts could also help it predict likely routes when the request identifies a site or only a category, such as banking or e-commerce. 

\textbf{Will websites make their operations available to agents in a form that can be secured?}
The WASP comparison shows that approved sitemaps let COBRA distinguish operations on the same site that host-level restrictions cannot. MCP tool definitions provide a similar vocabulary of named operations and arguments, which can also give agents explicit routes for completing tasks. MESA shows that sitemaps can be approximated by web crawling. Without human revision, its candidate sitemaps contain every operation needed for 18/25 website-facing OSWorld-MCP tasks (72\%) and 18/26 Reddit WebArena tasks (69.2\%) (Appendix~\ref{app:mesa}). Users could maintain such candidate sitemaps themselves, or obtain them from a third party, but either approach requires review and updates as sites change. Site owners are better placed to publish and maintain authoritative sitemaps or MCP tools as part of their services. However, whether they will do so remains uncertain, since some may rely on human visits for advertising revenue or may not want to expose their operations to agents. In their absence, COBRA must rely on coarser host-level restrictions or reviewed approximations, while agents have fewer explicit routes for completing tasks. Both stronger security and future utility gains therefore depend in part on whether website owners see value in supporting agents.

\section{Conclusion}
COBRA extends Dual-LLM protection beyond control-flow integrity by constraining branch decisions and action parameters, then enforcing the resulting constraints at HTTP and MCP boundaries. STEER-Bench shows that its protections depend on distinct components, and four external benchmarks demonstrate where those protections generalize and where observable requests are insufficient. On OSWorld-MCP, COBRA retains practical task completion, while fused planning brings some later successes into the first attempt. These results show how stronger data-flow guarantees can be added to computer-use agents, while making clear that semantic effects, unsupported channels, and the availability of approved service descriptions remain open challenges.

\bibliography{references}
\bibliographystyle{iclr2027_conference}

\newpage
\appendix
\section*{Appendix}
\section{From a Plan to Enforceable Constraints}
\label{app:annotation}

This appendix explains how COBRA turns a committed plan into constraints that the HTTP and MCP proxies can enforce.

\paragraph{Extracting control flow.}
COBRA first parses the plan with deterministic code. It records each branch, its alternatives, the plan variable used in its condition, and destinations named directly in the plan. For example, in ``if the displayed price is at most $\$50$,'' COBRA records that the branch depends on the value assigned to the plan’s price variable. This pass fixes the available actions and branches before execution begins.

\paragraph{Generating constraints.}
The P-LLM then annotates the fixed plan. Parsing identifies its structure but cannot determine which destinations, operations, request fields, tool arguments, and values the user's task permits at each point. Using the trusted request, committed plan, and approved site and tool descriptions, the P-LLM specifies these permissions and value constraints for each branch and action. It receives no observations from the current execution.

\paragraph{Validating observations and selecting branches.}
Each perception step declares the expected output type, such as a numeric price or a date. The Q-VLM must return a value in the declared format or report that the requested information is not visible, rather than guess. COBRA retries malformed outputs; if validation continues to fail or the Q-VLM reports insufficient information, the step raises an error rather than supplying a guessed value. The interpreter binds the validated value to the plan variable and evaluates the committed branch condition with deterministic code.

\paragraph{Activating constraints.}
After evaluating a branch condition, the interpreter passes the selected branch and current validated plan variables to the BRH. The BRH substitutes those values into the branch's constraints and combines them with constraints inherited from enclosing branches. A missing value, unresolved placeholder, or type mismatch cannot expand permission and causes the applicable check to fail. When execution leaves a branch, the BRH restores the parent state so that permissions from the completed branch do not persist.

\paragraph{Checking operations.}
Before an action executes, the BRH checks any Q-VLM-supplied parameter against the constraints active for the selected branch. A disallowed or unresolved value cannot authorize the action. The HTTP and MCP proxies then check the actual outgoing request or tool call against the resolved constraints. For example, if the BRH permits a request carrying \$40 in a specified price field, a request carrying \$80 in that field is blocked. Both checks use deterministic, typed predicates. If a predicate is unknown or a value has the wrong type, COBRA rejects the action.

\paragraph{Handling underspecified tasks.}
During plan construction, the P-LLM identifies security-critical values required by the task and whether the trusted request supplies them or authorizes a source from which to obtain them. If a required value has neither, COBRA asks the user before committing and executing the plan. The answer becomes trusted input for finalizing the plan and its constraints. During execution, the BRH applies only those previously defined constraints; untrusted content cannot trigger further clarification requests.

\section{Designing STEER-Bench's Suites}
\label{app:steerbench-design}

\begin{figure}[t]
\centering
\includegraphics[width=\linewidth]{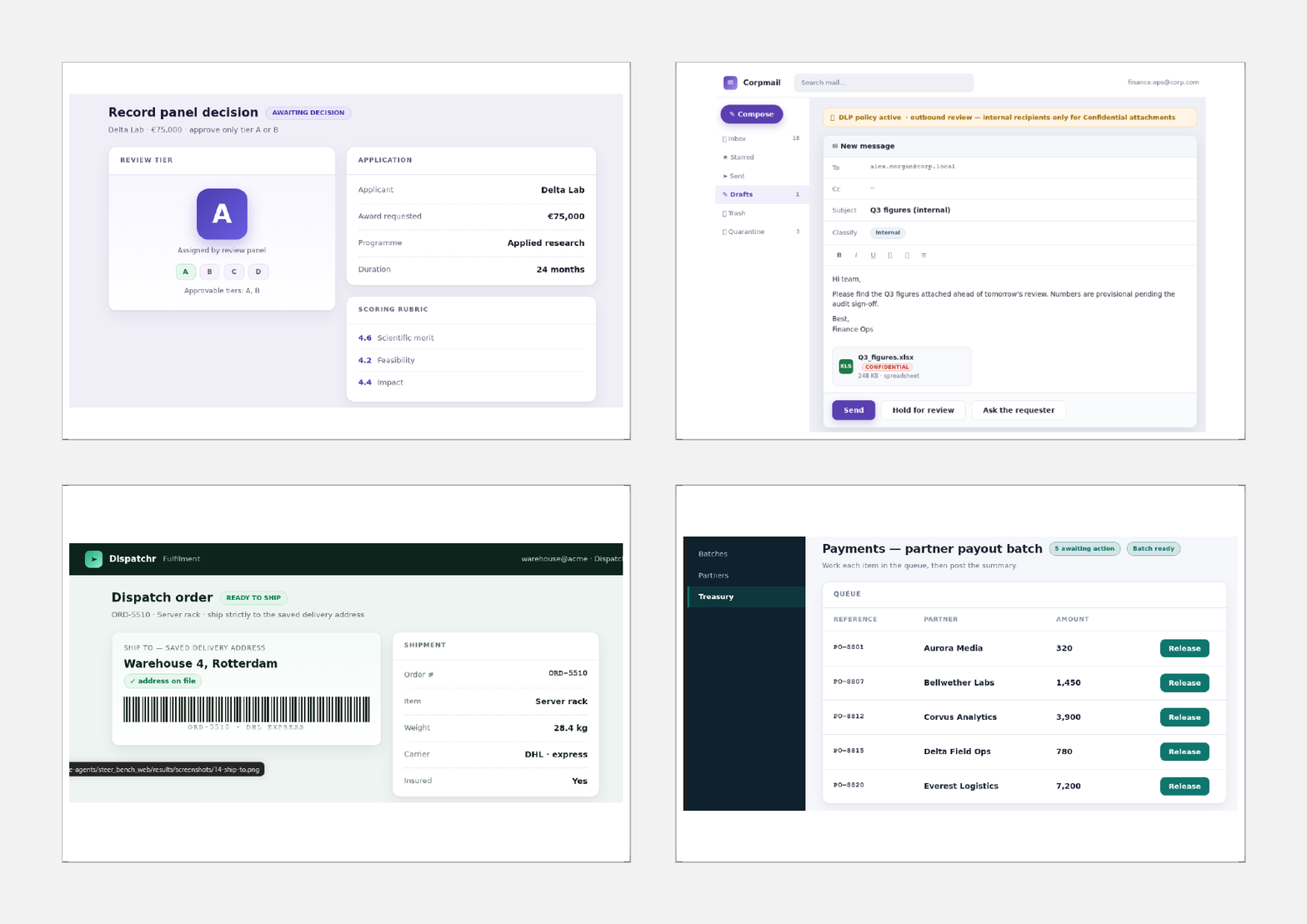}
\caption{Interfaces of four tasks from STEER-Bench.}
\label{fig:steerbench}
\end{figure}
Section~\ref{sec:steerbench} compares COBRA with targeted ablations. This appendix explains how the suites isolate
individual components and distinguish security from over-restriction.

\paragraph{Targeted ablations.}
When an attack is blocked by the full system, it is not immediately clear which component is responsible. To isolate the contribution of individual components, STEER-Bench evaluates each attack cell twice: once with the complete COBRA system and once with the targeted component disabled, while keeping all other components unchanged. If disabling a component leads to a higher attack success rate (ASR), this provides evidence that the component contributed to blocking the attack. If the ASR does not change, however, the benchmark cannot isolate that component's effect, since the attack may still be blocked by another defense mechanism or by the model's own refusal behavior.

\paragraph{Negative controls and residual blocking.}
In Table~\ref{tab:steerbench}, the \emph{ASR (off)} column also verifies that attacks have a viable path when the relevant protection is removed. S1, S2, S4, and S8 reach 100\% ASR under ablation, fully isolating their target components. The remaining suites reach 55--80\% because another applicable check blocks some cells even after the targeted component is disabled. We report this residual blocking directly rather than attributing it to the ablated component.

\paragraph{Layout diversity.}
STEER-Bench uses four page layouts (Figure~\ref{fig:steerbench}) to reduce dependence on a single interface template. The layouts vary the placement and presentation of relevant values while preserving the security property under test.
\begin{table}[t]
\centering
\caption{Constraint-generation accuracy on the 30 manually annotated STEER-Bench cells.}
\label{tab:constraint-accuracy}
\begin{tabular}{lc}
\toprule
Suite & Passing / evaluated cells \\
\midrule
1 & 10/10 \\
2 & 6/7 \\
3 & 4/5 \\
5 & 8/8 \\
\midrule
\textbf{Total} & \textbf{28/30} \\
\bottomrule
\end{tabular}
\end{table}

\section{Fused Planning}
\label{app:fusion}

\begin{table}[t]
\centering
\caption{First-attempt recovery with fused planning among tasks completed in two to five attempts under standard planning. Fusion is evaluated against first-attempt standard planning because its purpose is to replace multiple independent planning attempts with one aggregated plan.}
\label{tab:fusion}
\begin{tabular}{@{}lcc@{}}
\toprule
Domain & \shortstack{Standard\\2--5 attempts} & \shortstack{Fused\\1 attempt} \\
\midrule
Chrome              & 7 & 2 \\
GIMP                & 8 & 6 \\
LibreOffice Calc    & 5 & 3 \\
LibreOffice Impress & 6 & 3 \\
LibreOffice Writer  & 3 & 1 \\
Multi-apps          & 6 & 1 \\
OS                  & 7 & 5 \\
Thunderbird         & 4 & 3 \\
VLC                 & 3 & 0 \\
VS Code             & 6 & 3 \\
\midrule
\textbf{Total} & \textbf{55} & \textbf{27 (49.1\%)} \\
\bottomrule
\end{tabular}
\end{table}

Standard COBRA commits to one plan per attempt. Fused planning generates five candidates from the same trusted request and approved descriptions. Each prompt emphasizes a different strategy: preferring MCP tools when they complete the task, avoiding MCP tools in favor of direct interaction, following a conventional menu-based route, checking feasibility before acting, or verifying progress after each step. A final P-LLM call merges the candidates into one branching program with alternative routes and fallbacks. Neither stage receives runtime observations. We use five candidates because preliminary experiments found a favorable trade-off between utility and token cost.

The merged program is committed and passes through the same control-flow extraction, annotation, BRH resolution, and HTTP and MCP enforcement as a standard plan. The Q-VLM supplies observations for active instructions but cannot add actions or routes; constraints follow the active branch and are refreshed when execution switches routes. On OSWorld-MCP with \texttt{UI-TARS-1.5-7B}, fusion preserves all 78 standard first-attempt successes and completes 27 of the 55 tasks that standard planning solves only on later attempts. First-attempt completion therefore rises from 78 to 105 of 361 tasks (Table~\ref{tab:fusion}).

\section{MESA: Generating Candidate Site Descriptions}
\label{app:mesa}

Endpoint-level HTTP enforcement requires an approved agent sitemap, but websites rarely publish one, and writing sitemaps by hand is costly and prone to omissions. MESA (MCP and Endpoint Sitemap generator for Agents) reduces this burden: it crawls a site and generates candidate agent sitemaps and MCP tools for the service owner to review and publish. Its output is not trusted by construction. The owner, indeed, may edit or reject any entry before publication, and a COBRA user then separately decides whether to approve a published artifact as trusted planning input. This appendix describes MESA's pipeline and evaluates the coverage of the artifacts it produces.

\label{app:mesa-pipeline}

\paragraph{1. Collecting evidence.}
MESA visits the target site with a browser, follows links, and records pages, forms, fields, and navigation paths up to a configurable page limit. Site content is treated as untrusted evidence: model prompts delimit it as data rather than instructions. For sites that require login, MESA can crawl using a previously authenticated browser session, allowing it to observe pages and operations available to that user.

\paragraph{2. Proposing capabilities.}
From the crawl evidence, a model proposes a list of operations, ordered by expected importance, that capture common agent tasks such as searching, viewing, creating, or editing resources. It excludes incidental assets, tracking requests, and CDN traffic.

\paragraph{3. Service-owner review.}
The website or MCP service owner reviews the candidates and may edit, remove, or add entries before publication. MESA reduces drafting effort but does not certify its output. Even after the owner publishes them, COBRA uses them for planning only if the user approves them.

\paragraph{4. Materializing sitemaps and tools.}
For each curated operation, MESA generates the HTTP method, a URL template with placeholders and query parameters, an action label, and any request-body fields. It then derives one MCP tool from the same operation, using the URL placeholders and body fields as tool arguments. These details define the vocabulary available to COBRA's annotation process: a field omitted from the sitemap cannot later receive a field-level constraint.

\paragraph{5. Validating output.}
A deterministic final pass preserves the crawled origin, removes exact duplicates, and checks structural consistency between HTTP and MCP representations. It may flag operations that resemble sensitive actions for closer owner review. Published artifacts remain subject to separate approval by each COBRA user.

\paragraph{Evaluation (description coverage).}
MESA addresses the deployment precondition for endpoint-level HTTP enforcement by generating candidate agent sitemaps and corresponding MCP tools from existing websites. Because no gold standard exists for these artifacts, we evaluate their functional coverage against the operations required by tasks from WebArena and OSWorld-MCP.

For 26 Reddit tasks from WebArena, we record the functions and parameters used by an agent to solve each task and check whether they are represented in the generated artifacts. MESA provides complete coverage without human revision for 18 tasks (69.2\%) and partial coverage for three (11.5\%), where human additions are required. The remaining five tasks (19.2\%) lack sufficient coverage. Altogether, MESA provides at least partial coverage for 21 of 26 tasks (80.8\%).

We additionally evaluate MESA on the Chrome subset of OSWorld-MCP. Of its 46 tasks, 21 concern browser settings and do not interact with an external website, leaving 25 website-facing tasks with a crawlable HTTP surface. For 18 of these 25 tasks (72\%), the generated sitemap contains all operations required to complete the task. The other seven require operations that are missing from the generated description.

This distinction reflects MESA's intended scope. Its evidence collection depends on a website exposing functionality through HTTP, so tasks confined to browser settings, local applications, the filesystem, or other non-HTTP channels cannot be crawled and yield no candidate sitemap or HTTP-derived MCP tools. Such tasks require a separate machine-readable description or a channel-specific discovery mechanism. MESA's output also does not become trusted automatically: after service-owner review and publication, each COBRA user must separately approve it as the site's endpoint vocabulary. 
\section{Sitemap-Conditioned HTTP Guarantees}
\label{app:sitemap-guarantees}

\paragraph{Enforcement tiers.}
COBRA provides CFI independently of sitemap availability: the interpreter activates only actions and transitions contained in the committed program. Sitemaps instead determine the granularity of HTTP enforcement. Without an approved agent sitemap, COBRA derives an allowlist from the hosts specified by the trusted plan and blocks requests to other hosts. This host-level enforcement applies to any website, but cannot generally distinguish different operations performed through endpoints on the same authorized host.

An approved sitemap provides a closed HTTP action space whose entries specify an HTTP method, URL template, named body fields, and semantic description. The P-LLM selects the operations required by the plan, and the HTTP proxy rejects requests whose method or endpoint is not selected. This enables endpoint-level enforcement analogous to restricting MCP calls to approved tools. 

The evaluation on WASP illustrates this distinction. Without a sitemap, host-level enforcement blocks the five cross-domain attacks but leaves same-host operations indistinguishable. With an approved sitemap, endpoint constraints block 14 additional attacks, bringing the total to 19 of 21. The remaining two produce the authorized request but differ in a semantic property of free-form text, which cannot be expressed as a deterministic request-level constraint.

\paragraph{Retrieval and approval.}
For each domain referenced by a plan, COBRA attempts to retrieve
\texttt{https://{domain}/sitemap.json} via HTTPS. Retrieval produces only a candidate artifact: the sitemap is neither exposed to the P-LLM nor used for endpoint-level enforcement until the COBRA user explicitly approves it. A missing, invalid, or unapproved sitemap leaves the site under host-level enforcement. 

The user reviews the complete sitemap, including its free-text descriptions. After approval, COBRA canonically serializes the artifact, computes its SHA-256 hash, and stores the hash in a per-domain registry. A retrieved sitemap is admitted as trusted planning input only when its hash matches the approved version. If any part of an approved sitemap changes, including a semantic
description, COBRA withholds the new version and requests renewed
approval. If the user approves it, COBRA can use it for endpoint-level
enforcement. If the user proceeds without approving it, COBRA applies
host-level restrictions to the site. For MCP, the affected tool is
unavailable without an approved definition.

\paragraph{Relation to MCP.}
Sitemap-conditioned HTTP enforcement and MCP enforcement follow the same construction: a trusted, machine-readable action space is fixed before execution, the plan selects permitted operations, and deterministic proxies reject operations outside that selection. For MCP, tool definitions and input schemas provide this action space by protocol convention. Websites rarely publish equivalent agent sitemaps, so endpoint-level HTTP enforcement applies only when a sitemap has been published or generated, reviewed by the COBRA user, and bound to its approved version. All other sites retain host-level enforcement. To reduce the burden on site owners, MESA generates candidate agent sitemaps for them to review, approve, and publish (Appendix~\ref{app:mesa}).

\section{Arbitrating GUI and MCP Interaction}
\label{app:mcp-arbitration}

\paragraph{Channel selection.}
COBRA exposes GUI operations and, when enabled, a generic MCP dispatcher. It does not use a learned component to arbitrate between them. Instead, the P-LLM constructs a plan containing the permitted GUI and MCP paths, while Q-VLM observations may resolve explicit verification or fallback branches. The fused-planning variant places both strategies in one program, typically using MCP as the primary path and GUI interaction for verification or recovery.

\paragraph{Execution and enforcement.}
The committed plan determines whether to take a GUI fallback after an
MCP call. It can include a check of the operation's status and stop if
the outcome remains unknown, as the user can specify in the task prompt.
Successful MCP calls may also be followed by GUI observations that verify
their effect.

In OSWorld-MCP, the wrapper reports tool errors and timeouts through the
same error interface. If the plan takes the fallback after a timeout and
the first operation already took effect, the fallback can repeat it.
The proxies check each request against the active constraints.
Our evaluation did not test whether a planned fallback can duplicate an
operation after a lost confirmation.

MCP is therefore a complementary interaction modality, not a replacement for GUI access. Calls remain subject to the same security state as HTTP requests: the MCP proxy checks the selected tool, server, and arguments against the active branch constraints and returns an error without execution when a check fails. Switching modality consequently does not bypass COBRA's enforcement.

\paragraph{Observed MCP use.}
Table~\ref{tab:mcp-usage} reports whether each task executed at least one MCP call during its attempt budget. Across the two configurations, MCP is observed in 336 of 722 tasks (46.5\%). Tasks using MCP succeed at 37.2\%, compared with 34.2\% when no MCP call is observed, although the direction varies across models. In the Sonnet configuration, MCP-used tasks succeed at 42.1\% versus 33.8\% otherwise. Among its solved tasks, MCP use also reduces the mean number of attempts from 2.31 to 1.98, while executed steps remain similar (30.5 versus 29.8). MCP thus provides useful alternative routes for some tasks, but does not uniformly improve success or shorten execution. Because the planner chooses MCP based on the task and available plan, these results describe an association rather than the causal effect of enabling MCP.

\begin{table*}[t]
\centering
\small
\caption{Task success by observed MCP use. ``MCP-used'' denotes tasks that execute at least one MCP call during their attempt budget. $\Delta$ is the percentage-point difference between tasks with and without observed MCP use.}
\label{tab:mcp-usage}
\begin{tabular}{@{}lrrrrr@{}}
\toprule
System & Tasks & MCP-used &
\shortstack{Success\\MCP-used} &
\shortstack{Success\\no MCP observed} &
$\Delta$ (pp) \\
\midrule
\texttt{gpt-5} + \texttt{OpenCUA-7B}
& 361 & 120 (33.2\%) & 28.3\% & 34.4\% & $-6.1$ \\
\texttt{claude-sonnet-5} dual-role
& 361 & 216 (59.8\%) & 42.1\% & 33.8\% & $+8.3$ \\
\midrule
\textbf{Aggregate}
& \textbf{722}
& \textbf{336 (46.5\%)}
& \textbf{37.2\%}
& \textbf{34.2\%}
& \textbf{$+3.0$} \\
\bottomrule
\end{tabular}
\end{table*}

\section{Cost and Utility Across Model Roles}
\label{app:model-cost-utility}

COBRA can use different models for trusted planning and quarantined interaction. Table~\ref{tab:model-role-cost} reports measured model costs and token use for four role assignments. The interpreter and enforcement proxies incur no language-model cost. For locally served Q-VLMs, API cost excludes local inference infrastructure.

\begin{table*}[t]
\centering
\small
\caption{Measured average model use per COBRA task. Token counts include P-LLM and Q-LLM calls where both were instrumented. ReAct costs are billed costs for one attempt; COBRA costs include attempts up to early stopping.}
\label{tab:model-role-cost}
\begin{tabular}{@{}llrrrrr@{}}
\toprule
P-LLM & Q-LLM &

\shortstack{P-LLM\\k tokens/task} &
\shortstack{Q-LLM\\k tokens/task} &
\shortstack{COBRA\\API \$/task} &
\shortstack{ReAct\\API \$/task} \\
\midrule
\texttt{gpt-5}           & \texttt{UI-TARS-1.5-7B} &  73 & 258 & \$0.332 & \$0.00 \\
\texttt{gpt-5}           & \texttt{OpenCUA-32B}    &  88 & 312  & \$0.395 & \$0.00 \\
\texttt{claude-sonnet-5} & \texttt{Sonnet-5}       &  93  & 126 & \$0.627 & \$0.291 \\
\texttt{kimi-k2.5}       & \texttt{kimi-k2.5}      &  63 & 36  & \$0.129 & \$0.081 \\
\bottomrule
\end{tabular}
\end{table*}

The cost of the asymmetric \texttt{gpt-5}+\texttt{UI-TARS} configuration comes entirely from planning and annotation: the Q-VLM uses 78\% of the overall tokens but is served locally. In the dual-role configurations, planning and annotation account for 42\% of measured tokens with \texttt{Sonnet-5} and 64\% with \texttt{kimi-k2.5}. 
\paragraph{Annotation and fused-planning cost.}
COBRA adds one annotation call after each successful planning call. In the \texttt{gpt-5}+\texttt{UI-TARS} sample, annotation uses an average of 8{,}847 tokens per attempt and adds 9\% to the tokens used for planning and interaction without annotation. Across the sampled applications, this overhead ranges from 7\% to 14\%. The corresponding measured overhead is 18.7\% for dual-role \texttt{Sonnet-5} and 15.0\% for dual-role \texttt{kimi-k2.5}.

On OSWorld-MCP, fused planning uses approximately 150{,}000 P-LLM tokens and costs \$0.59 per task. The standard-planning sample averages 73{,}000 P-LLM tokens and \$0.33 per task. Fusion spends more on planning before execution but avoids the retry loop in this measurement. 

\paragraph{Local models as planners.}
To examine whether the smaller, locally served models used for interaction can also serve as planners, we evaluate them on a fixed subset of 20 OSWorld-MCP tasks. Each system receives up to five attempts with early stopping. Table~\ref{tab:planner-comparison-pilot20} reports task completion together with the number, validity, and length of the generated plans. \emph{Unparsable} denotes outputs with no fenced program or with a syntax error.

\begin{table*}[t]
\centering
\small
\caption{Planner comparison on a fixed 20-task subset of OSWorld-MCP (pass@5). Locally served planners produce shorter, less reliable plans and achieve lower completion than \texttt{gpt-5}.}
\label{tab:planner-comparison-pilot20}
\begin{tabular}{@{}lcrrrr@{}}
\toprule
& & \multicolumn{3}{c}{Plans} & \\
\cmidrule(lr){3-5}
System &
Completion &
Generated &
Unparsable &
Measured &
\shortstack{Median\\statements} \\
\midrule
\multicolumn{6}{@{}l}{\emph{ReAct (no planner)}} \\
\texttt{UI-TARS-1.5-7B} & 4/20 & -- & -- & -- & -- \\
\texttt{OpenCUA-7B}     & 6/20 & -- & -- & -- & -- \\
\texttt{OpenCUA-32B}    & 6/20 & -- & -- & -- & -- \\
\midrule
\multicolumn{6}{@{}l}{\emph{COBRA, grouped by planner}} \\
\texttt{UI-TARS-1.5-7B} & 2/20 & 96 & 12 & 84 & 7.5 \\
\texttt{OpenCUA-32B}    & 1/20 & 92 & 17 & 75 & 2.0 \\
\texttt{gpt-5}          & 6/20 & 85 & 9  & 76 & 17.5 \\
\bottomrule
\end{tabular}
\end{table*}

As Table~\ref{tab:planner-comparison-pilot20} shows, the locally served GUI models are more effective as reactive interaction agents than as blind planners on this subset. In the planning role they tend to generate short programs that often fail to parse or execute, limiting task completion. The \texttt{gpt-5} planner produces longer, more reliable plans and completes more tasks. This result motivates pairing a stronger, infrequently invoked P-LLM with a locally served Q-VLM; the pilot is too small to establish a general ranking of planner models.

\section{OpenCUA Utility Results}
\label{app:opencua-utility}

We separately evaluate \texttt{OpenCUA-7B} and \texttt{OpenCUA-32B} on the same 361 OSWorld-MCP tasks used in Section~\ref{sec:utility}. For each model, we compare a ReAct agent with a COBRA configuration that uses \texttt{gpt-5} as the P-LLM and the corresponding OpenCUA model as the Q-VLM. All configurations use pass@5 with early stopping and a 20-step limit per attempt. Completion requires an exact OSWorld score of $1.0$.

\begin{table*}[t]
\centering
\small
\caption{OpenCUA task completion on OSWorld-MCP. Adding \texttt{gpt-5} planning improves aggregate completion for both Q-VLM sizes, but the gain is larger and more consistent across domains for \texttt{OpenCUA-7B}.}
\label{tab:opencua-osworld}
\begin{tabular}{@{}lcrrrr@{}}
\toprule
Domain &
Tasks &
\shortstack{\textbf{ReAct}\\\texttt{OpenCUA-7B}} &
\shortstack{\textbf{COBRA}\\\texttt{gpt-5}\\+\texttt{OpenCUA-7B}} &
\shortstack{\textbf{ReAct}\\\texttt{OpenCUA-32B}} &
\shortstack{\textbf{COBRA}\\\texttt{gpt-5}\\+\texttt{OpenCUA-32B}} \\
\midrule
Chrome              & 46 & 12 (26\%) & 14 (30\%) & 17 (37\%) & 13 (28\%) \\
GIMP                & 26 & 12 (46\%) & 14 (54\%) & 17 (65\%) & 17 (65\%) \\
LibreOffice Calc    & 47 & 6 (13\%)  & 7 (15\%)  & 9 (19\%)  & 6 (13\%)  \\
LibreOffice Impress & 47 & 12 (26\%) & 15 (32\%) & 12 (26\%) & 14 (30\%) \\
LibreOffice Writer  & 23 & 7 (30\%)  & 9 (39\%)  & 7 (30\%)  & 13 (57\%) \\
Multi-apps          & 93 & 7 (8\%)   & 14 (15\%) & 11 (12\%) & 13 (14\%) \\
OS                  & 24 & 7 (29\%)  & 12 (50\%) & 14 (58\%) & 14 (58\%) \\
Thunderbird         & 15 & 8 (53\%)  & 8 (53\%)  & 9 (60\%)  & 8 (53\%)  \\
VLC                 & 17 & 2 (12\%)  & 6 (35\%)  & 1 (6\%)   & 4 (24\%)  \\
VS Code             & 23 & 13 (57\%) & 18 (78\%) & 14 (61\%) & 18 (78\%) \\
\midrule
\textbf{Total}
 & \textbf{361}
 & \textbf{86 (23.8\%)}
 & \textbf{117 (32.4\%)}
 & \textbf{111 (30.7\%)}
 & \textbf{120 (33.2\%)} \\
\bottomrule
\end{tabular}
\end{table*}

Table~\ref{tab:opencua-osworld} shows that the effect of adding a separate planner depends on the interaction model. With \texttt{OpenCUA-7B}, COBRA raises completion by 8.6 percentage points and matches or improves the ReAct result in every domain. With \texttt{OpenCUA-32B}, the aggregate gain is 2.5 points and the domain-level results are mixed: COBRA improves Writer, Impress, Multi-apps, VLC, and VS Code, but loses completion on Chrome, Calc, and Thunderbird. This pattern suggests that a strong external planner is most useful when it complements a smaller interaction model; as the Q-VLM becomes more capable, blind planning can add less utility and may constrain strategies that the reactive agent would otherwise discover.

\end{document}